\documentstyle[12pt]{preprint}

\pubdate{July 1991}
\pubnumber{EFI-91-41}
\title{A $\kappa$-Symmetry Calculus for Superparticles}
\author
{Jerome P. Gauntlett\thanks
{E-mail: Jerome@yukawa.uchicago.edu}
\address{\efi}}
\abstract{
We develop a $\kappa$-symmetry
calculus for the d=2 and d=3, N=2
massive superparticles, which enables us to construct
higher order $\kappa$-invariant actions.
The method relies on a reformulation of these
models as supersymmetric sigma models that are
invariant under local worldline superconformal
transformations. We show that the
$\kappa$-symmetry is embedded in
the superconformal symmetry so that a calculus
for the $\kappa$-symmetry is equivalent to
a tensor calculus for the latter. We develop
such a calculus without the introduction of a
worldline supergravity multiplet.
}

\begin{document}
\maketitle

\def \p2{(-\pi^2)^{1/2}}

{\bf 1.} One of the major obstacles in the
covariant quantisation of the Green-Schwarz
superstring is a lack of understanding of
the local fermionic ``$\kappa$-symmetry". A more
geometric world-sheet interpretation of this
symmetry is desirable in order to make further progress.
A useful strategy to develop a deeper understanding of
$\kappa$-symmetry is to try to construct a
tensor calculus allowing one to write down higher
order actions (i.e. containing the extrinsic curvature
of the worldsheet) that are
invariant under $\kappa$-symmetry.

There are also more specific reasons for wanting to
construct such a $\kappa$-symmetry
calculus. Supersymmetric extended
objects are generalisations of the Green-Schwarz
superstring and are invariant under $\kappa$-symmetry.
It is now well established that these actions have
the physical interpretation as the leading term
in the effective action describing the low energy
dynamics of topological defects of supersymmetric
field theories [1-8]. Higher order
$\kappa$-invariant actions thus correspond to
possible correction terms and may
play an important role in processes such as
radiation from cusps and kinks.
Furthermore, in the quantum theory of supersymmetric
extended objects, these terms will correspond to
possible counter terms and/or covariant regulators.

There has been previous work on the construction of
higher order $\kappa$-invariant actions.
Actions for superstrings,
quadratic in
the extrinsic curvature of the worldsheet,
were constructed using a direct approach in [9]
and the invariance under the $\kappa$-symmetry
was partially checked. By a dimensional reduction
of the action proposed in [9], an action for a d=2
superparticle quadratic in the extrinsic curvature
of the worldline was proposed in [10] and the invariance
under the $\kappa$-symmetry was fully checked.
In [10,11,7] (see also [8]) a systematic
method was developed for writing down
higher order actions, by exploiting
the fact that in the
``physical gauge"
supersymmetric extended objects exhibit a partial
breaking of rigid supersymmetry [1-7].
Although this is a
systematic method it only delivers actions in
gauge-fixed form and thus the underlying
$\kappa$-symmetry is
obscured.

In this paper we will develop a $\kappa$-symmetry
calculus for the simplest supersymmetric extended
objects, massive superparticles. Specifically,
we will develop the calculus for the d=2 and the d=3, N=2
massive superparticles.
Following the work on
massless superparticles by Sorokin, Tkach and
Volkov (STV) [12] and others [13],
our method relies on a reformulation of these
actions as N=1 and N=2 supersymmetric sigma models
that are invariant under local worldline
superconformal transformations.
We show that the $\kappa$-symmetry is
embedded in
the superconformal symmetry, so that developing
a tensor calculus for the latter will
provide
us with the sought after $\kappa$-symmetry
calculus. We illustrate the formalism by
constructing $\kappa$-invariant actions
quadratic in the extrinsic curvature of the
worldline, for the two cases.

When this work was near completion we received
two preprints by Ivanov and Kapustnikov [14,15] who discuss
similar issues. They construct a $\kappa$-symmetry
calculus for the d=2 massive superparticle, obtaining
similar results to those presented here
and they also produce some preliminary
results for the d=4 superstring.

{\bf 2.} In [12] STV reformulated the massless
Brink-Schwarz superparticle as an N=1 supersymmetric
sigma model with flat d-dimensional superspace
as the target space, for d=3, 4, 6 and 10. Remarkably,
the STV action is invariant under local N=1
superconformal transformations without the
introduction of a worldine supergravity
multiplet. They further showed that
the odd sector of these
gauge transformations correspond to one of
the components of the $\kappa$-symmetry
of the Brink-Schwarz superparticle.
Since for d=3, 4, 6 and 10 the $\kappa$-symmetry
has 1, 2, 4 and 8 independent components,
the STV and the Brink-Schwarz
actions are
actually equivalent for d=3.
By considering a dimensional reduction of the
STV action in d=3 we are led to the following
new reformulation
of the d=2 massive superparticle as an
N=1 supersymmetric sigma model.
This will then allow us to construct a
$\kappa$-symmetry calculus.

We first introduce N=1 worldline superspace
with co-ordinates $z^M=(t,\eta)$ and a flat d=2 target
superspace with co-ordinates
$(X^\mu(z),\Theta_\alpha(z))$,
where $X^\mu$ and $\Theta_\alpha$ are
worldline scalar superfields transforming as
a vector and a real two component Majorana
spinor of the d=2 superPoincar\'e group,
respectively, subject to the constraint
$$
DX^\mu+iD\bar\Theta\Gamma^\mu\Theta=0\qquad.\eqno(1)
$$
The new action for the d=2 massive
superparticle is then given by
$$
S_1=-m\int d^2z D\bar\Theta\Gamma_3\Theta\eqno(2)
$$
where
$D\equiv\partial_\eta-i\eta\partial_t$,
$\Gamma_3\equiv\Gamma^0\Gamma^1$
and our spacetime has signature $(-,+)$.

The constraint and action are both invariant
under rigid $d=2$ superPoincar\'e transformations,
$$
\eqalign{\delta X^\mu&=i\bar\rho\Gamma^\mu\Theta\cr
\delta \Theta&=\rho\qquad,}\eqno(3)
$$
where $\rho$ is a constant spacetime spinor,
but we note that the Lagrangian transforms
as a kind of Wess-Zumino term.

The action is also clearly invariant under
rigid supersymmetry transformations of
the worldline. What
is less obvious is that it is actually
invariant under a restricted
group of worldline superdiffeomorphisms and
superWeyl transformations,
one dimensional local superconformal
transformations, whose
infinitesimal form is given by
$$
\delta z^M=A\dot z^M+{i\over 2}DADz^M\qquad,\eqno(4)
$$
where $A(z)$ is a real scalar superfield.
This can be seen by observing that
$$
\delta D\equiv D' -D=-{\dot A\over 2}D\qquad,\eqno(5)
$$
$$
\delta d^2z\equiv d^2z'-d^2z={\dot A\over 2}d^2z\eqno(6)
$$
and that the scalar fields are invariant
(e.g. $\delta X^\mu\equiv X'^{\mu}(z')-X^\mu(z)=0$).
We note that
the transformations (4) can in fact be determined
by demanding that the
flat superspace covariant derivative
transforms homogeneously as in (5).

To construct the component form of the action
we  first deduce from the
constraint the following component
expansions for the superfields
$$
\eqalign{X^\mu&=x^\mu-i\eta\bar\lambda\Gamma^\mu\theta\cr
\Theta_\alpha&=\theta_\alpha
+\eta\lambda_\alpha\cr}\eqno(7)
$$
and in addition that
$$
\pi^\mu=\bar\lambda\Gamma^\mu\lambda
\qquad\Rightarrow\qquad \p2=\bar
\lambda\Gamma_3\lambda\qquad,
\eqno(8)
$$
where
$\pi^\mu\equiv \dot x^\mu-i\bar
\theta\Gamma^\mu\dot\theta$.
Actually there is a sign ambiguity in the second
relation of (8) but we work with one sign
for definiteness; all of the
following analysis could be
repeated for the other choice.
After performing the $\eta$-integration
in (2) we obtain
$$
S_1=-m\int dt\{(-\pi^2)^{1/2}
+i\bar\theta\Gamma_3\dot\theta\}\qquad,\eqno(9)
$$
which is the standard
d=2 massive superparticle action [4].

The local worldline supersymmetry
transformations (4) induce
transformations on a scalar field
$\Phi$ via
$$
\eqalign{\tilde\delta\Phi&\equiv \Phi'(z)-\Phi(z)\cr
&=-\delta z^M\partial_M\Phi\qquad.\cr}\eqno(10)
$$
Setting $A|=-a(t)$ and $DA|=2i\epsilon(t)$
(where a vertical bar means setting $\eta=0$),
we obtain the following
transformations on the component fields
$$
\eqalign{
\tilde\delta\theta&=a\dot\theta+\epsilon\lambda\cr
\tilde\delta x^\mu&=a\dot x^\mu+i\bar\theta
\Gamma^\mu(\epsilon\lambda)\qquad.\cr}\eqno(11)
$$
The transformations with parameter $a$
are the time reparametrisations of
the action (9). To identify the
$\kappa$-symmetry we
choose $\epsilon=-{\bar\lambda
\Gamma_3\kappa\over \p2}$, where
$\kappa(t)$ is a real two component
spacetime spinor, so that
$$
\eqalign{\tilde\delta_\kappa x^\mu & =i
\bar\theta\Gamma^\mu (\tilde\delta_\kappa\theta)\cr
\tilde\delta_\kappa\theta &={1\over 2}\left\{1+\Gamma
\right\}\kappa\cr}\eqno(12)
$$
where
$$
\Gamma\equiv
{\pi^\mu \Gamma_\mu \Gamma_3\over \p2}\qquad.\eqno(13)
$$
These are  precisely the $\kappa$-symmetry
transformations of the action (9) [4].
Thus, the $\kappa$-symmetry transformations
are equivalent to the odd sector of the superconformal
transformations and hence the action (2) is
indeed an equivalent reformulation
of the superparticle action
(9).

The construction of a $\kappa$-symmetry
calculus for the d=2 massive
superparticle is now equivalent
to finding a calculus for the local
superconformal transformations (4).
We will do this without
the introduction of a worldline supergravity
multiplet. We first introduce the density
$$
E\equiv \sqrt{D\bar\Theta\Gamma_3D\Theta}\qquad,\eqno(14)
$$
which transforms as
$$
\delta E= -{\dot A\over 2}E\eqno(15)
$$
and hence the measure $d^2zE$ is invariant.
This density can also be used to construct
the covariant derivative
$$
\nabla=E^{-1}D\qquad,\eqno(16)
$$
so that if $\Phi$ is a scalar
superfield then so is $\nabla\Phi$.
Taking the constraint (1) into account,
the superconformally invariant
and hence $\kappa$-invariant
actions of physical interest are given by
$$
S=\int d^2zE{\cal L}(\Theta,\nabla
\Theta,\nabla^2\Theta,...)\eqno(17)
$$
where ${\cal L}$ is a superPoincar\'e invariant,
Grassmann odd, worldine scalar.

As an illustration of this formalism we
consider the following action
$$
S_2=-{4\over m}\int d^2zE\nabla^3\bar
\Theta\Gamma_3\nabla^2\Theta\qquad.\eqno(18)
$$
After performing the $\eta$-integration we obtain
$$
\eqalign{
S_2={4\over m}\int dt (\bar\lambda\Gamma_3
\lambda)^{-6}\big[&-(\dot{\bar
\lambda}\Gamma_3\lambda)^2+
(\bar\lambda\Gamma_3\lambda)
(\dot{\bar\lambda}\Gamma_3\dot\lambda)+
3i(\bar\lambda\Gamma_3\dot\theta)
(\dot{\bar\lambda}\Gamma_3\dot\theta)\cr
 & + i(\bar\lambda\Gamma_3\lambda)
(\dot{\bar\theta}\Gamma_3\ddot\theta)+
i(\bar\lambda\Gamma_3\ddot\theta)
(\bar\lambda\Gamma_3\dot\theta)\big]\cr}\eqno(19)
$$
and using the constraints (8) and some
Fierz rearrangements we can rewrite this as
$$
S=m^{-1}\int dt\left\{-{{\cal B}^2\over \p2}
+{i{\bar F}\pi^\mu\Gamma_\mu\dot
F\over (-\pi^2)^{3/2}}\right\}
\eqno(20)
$$
where
$$
{\cal B}_\mu\equiv
{d\over dt}\left({\pi^\mu\over{\p2}}\right)
 +{1\over 2}{i\bar G\Gamma_\mu F\over \p2}\eqno(21)
$$
and
$$
F\equiv\Gamma_3(1-\Gamma)\dot\theta,
\qquad G\equiv\Gamma_3(1+\Gamma)\dot
\theta\qquad.\eqno(22)
$$
This is precisely the $\kappa$-invariant action
quadratic in the extrinsic curvature of the
worldline that was constructed
in [10].

{\bf 3.} We now generalise these results
to the case of the N=2 massive superparticle in d=3.
By a dimensional reduction of the
reformulation of the d=4 massless superparticle presented
in [13] (differing slightly
from that proposed by STV in [12]),
we obtain the following reformulation of the
d=3, N=2 massive superparticle as an N=2
worldline supersymmetric sigma model. The $\kappa$-symmetry
has two independent components and we
show that it is equivalent to the
odd sector of the local
superconformal symmetry.

We begin by introducing N=2 worldline
superspace with co-ordinates
$z^M=(t,\eta,\bar\eta)$ and a d=3, N=2
flat target superspace with co-ordinates
$(X^\mu(z),\Theta^I_\alpha(z))$,
where $X^\mu$ and $\Theta^I_\alpha$
transform as a vector and two two
component real spinors of the d=3 spacetime
superPoincar\'e group and as
worldline scalar superfields.
Defining
$$
\Psi_\alpha\equiv \Theta^1_\alpha
+i\Theta^2_\alpha,\qquad
X^\mu_L\equiv X^\mu+{i\over 2}\bar
\Psi\Gamma^\mu\Psi\qquad\eqno(23)
$$
and
$$
D\equiv\partial_\eta+i\bar\eta\partial_t,
\qquad
\bar D \equiv D^{*}=-\partial_{\bar\eta}
-i\eta\partial_t}\qquad,\eqno(24)
$$
the superfields are subject to the constraints
$$
\bar D X^\mu_L=0,\qquad
\bar D \Psi=0\qquad {\rm and}\qquad c.c.\eqno(25)
$$
where $c.c.$ means ``complex conjugate".

 From these constraints we can determine
the general component expansions
$$
\eqalign{
X^\mu&=x^\mu-{i\over 2}\eta\bar\psi\Gamma^
\mu\lambda-{i\over 2}\bar\eta\bar
\lambda\Gamma^\mu\psi
-{1\over 2}\eta\bar\eta({\dot{\bar\psi}}
\Gamma^\mu\psi+\bar\psi\Gamma^\mu\dot\psi)\cr
\Psi_\alpha&=\psi_\alpha+\eta\lambda_\alpha
+i\eta\bar\eta\dot\psi_\alpha,
\qquad {\rm and}\qquad c.c.\cr}\eqno(26)
$$
and the identities
$$
\pi^\mu=-{1\over 2}\bar \lambda\Gamma^\mu
\lambda\qquad\Rightarrow\qquad \p2={i\over 2}
\bar \lambda\lambda\qquad,\eqno(27)
$$
where $\pi^\mu=\dot x^\mu-i\theta^I\Gamma^
\mu\dot\theta^I$ and $\psi\equiv\theta^1+i\theta^2$.
Again there is a sign ambiguity in the
last relation and we have chosen a definite sign.

The new action for the d=3 N=2 massive
superparticle is given by
$$
S_3=-{im\over 2}\int d^3z \bar\Psi\Psi\eqno(28)
$$
and after integration of the Grassmann
co-ordinates leads to
the standard form of the superparticle action [7]
$$
S_3=-m\int dt\left[\p2+i\bar\theta^I\dot
\theta^J\epsilon^{IJ}\right]\qquad.\eqno(29)
$$

The constraints and the action are clearly
invariant under the N=2, d=3 spacetime
supersymmetry transformations
$$
\eqalign{
\delta X^\mu&=i\bar\rho^I\Gamma^\mu\Theta^I\cr
\delta\Theta^I&=\rho^I\qquad,\cr}\eqno(30)
$$
where $\rho^I$ are two constant anticommuting
spacetime spinors. We note that
just as
for the d=2 superparticle, the Lagrangian
is again a kind of Wess-Zumino term.

The action is also
invariant under the local N=2 superconformal
transformations of the worldline
$$
\delta z^M= A\dot z^M+{i\over 2}\bar D ADz^M
+{i\over 2}DA\bar Dz^M\qquad.\eqno(31)
$$
This can be seen by observing that the
measure is invariant
$$
\delta d^3z\equiv d^3z'-d^3z=0\eqno(32)
$$
and that the scalar fields are invariant.

The transformations (31) induce transformations
on scalar fields as in (10).
Choosing $A|=-a(t)$, $DA|=-2i\bar\epsilon(t)$
and $\bar D A|=2i\epsilon(t)$,
we deduce the following transformations on
the component fields:
$$
\eqalign{
\tilde\delta x^\mu&=a\dot x^\mu+{i\over 2}
\bar\psi\Gamma^\mu\epsilon\lambda-{i\over 2}
\bar\epsilon
\bar\lambda\Gamma^\mu\psi\cr
\tilde\delta \psi&=a\dot\psi+\epsilon\lambda
\qquad {\rm and}\qquad c.c.\cr}\eqno(33)
$$
The transformations with parameter $a(t)$
are the time reparametrisations
of the action (29). Choosing
$\epsilon={\bar\lambda\kappa\over
{\bar\lambda\lambda}}$, where $\kappa\equiv
\kappa^1+i\kappa^2$ is a complex
two component spinor, we find
$$
\eqalign{
\tilde\delta x^\mu&=i\bar\theta^I\Gamma^
\mu\tilde\delta\theta^I\cr
\tilde\delta\theta^I&={1\over 2}\left[\delta^{IJ}+
{\pi^\mu\Gamma_\mu\over\p2}\epsilon^{IJ}
\right]\kappa^J\qquad,\cr}\eqno(34)
$$
which are the standard $\kappa$-symmetry
transformations of the action (29)
(see [7]). Thus the time
reparametrisations and the $\kappa$-symmetry
are embedded in the N=2 local
superconformal transformations
. We note that
the components
$x^\mu$ and $\theta_\alpha$ are inert
under the local $U(1)$
transformations (corresponding to the component
$\bar D D A|$).

The construction of a $\kappa$-symmetry
calculus is now equivalent to finding
a method for constructing actions invariant
under the local N=2 superconformal transformations.
Although the measure $d^3z$ is invariant,
the covariant derivatives transform as
$$
\eqalign{
\delta D\equiv D'-D&=-{i\over 2}D\bar D AD\cr
\delta\bar D\equiv\bar D'-\bar D&
=-{i\over 2}\bar D DA\bar D\qquad.\cr}\eqno(35)
$$
We note that demanding that the flat covariant
derivatives transform homogeneously
is in fact enough to determine the form of the
transformations (31).
To construct the covariant derivatives
we first introduce the density
$$
E\equiv  \sqrt {i\bar D\bar\Psi D\Psi}\eqno(36)
$$
which transforms as
$$
\delta E=-{\dot A\over 2}E\qquad.\eqno(37)
$$
We define covariant derivatives of
scalar superfields as
$$
\nabla\Phi\equiv E^{-1}D\Phi,\qquad
\bar\nabla\Phi\equiv E^{-1}\bar D\Phi\eqno(38)
$$
with transformation laws
$$
\delta\left(\nabla \Phi\right)
=iL\nabla\Phi,\qquad \delta\left(\bar
\nabla\Phi\right)=-iL\bar\nabla\Phi\qquad,\eqno(39)
$$
where
$$
L=\bar L=-{1\over 4}[D,\bar D]_{\_}A\qquad.\eqno(40)
$$
Since the tangent space group for N=2
worldline supergravity is $U(1)$ (see [16] for
further details), we see that
the covariant derivatives $\nabla$
and $\bar\nabla$ carry U(1) charge
$+1$ and $-1$, respectively.
To construct higher order covariant
derivatives we need to generalise
the action of the covariant derivatives
on scalar superfields to objects
that transform with $U(1)$ charge n:
$$
\delta\Omega=niL\Omega\qquad.\eqno(41)
$$
Specifically, we define
$$
\nabla\Omega=E^{-1}(D+nE^{-1}DE)\Omega,
\qquad  \bar\nabla\Omega=E^{-1}
(\bar D-nE^{-1}\bar D E)\Omega\eqno(42)
$$
so that
$$
\delta(\nabla\Omega)=(n+1)iL\nabla
\Omega,\qquad  \delta(\bar\nabla\Omega)
=(n-1)iL\bar\nabla\Omega\qquad.\eqno(43)
$$
In particular, the covariant time
derivative acting on a scalar field is given by
$$
-2i\nabla_0\Phi\equiv(\nabla\bar\nabla
+\bar\nabla\nabla)\Phi\eqno(44)
$$
where
$$
\nabla_0\Phi=E^{-2}(\partial_t-iE^
{-1}DE\bar D -iE^{-1} \bar D E D)
\Phi\qquad.\eqno(45)
$$
We now have all the tools to construct
superconformally invariant and hence
$\kappa$-invariant actions.
The actions of physical interest are of the form
$$
S=\int d^3z {\cal L}(\Psi,\bar\Psi,
\nabla\Psi,\bar\nabla\bar\Psi,
\nabla_0\Psi,...)\eqno(46)
$$
where ${\cal L}$ is a superPoincar\'e
invariant, Grassmann even,
worldline scalar functional.

We illustrate this formalism by
constructing a $\kappa$-invariant
action that is quadratic in the extrinsic
curvature of the worldline:
$$
S_4=-8im^{-1}\int d^3z\nabla_0\bar
\Psi\nabla_0\Psi\qquad.\eqno(47)
$$
After carrying out the Grassmann
integration we obtain the component
form of the action whose bosonic piece
is given by
$$
S_4^{bosonic}=
m^{-1}\int dt \left[{B^2\over
\p2\right]_{bosonic}\qquad,\eqno(48)
$$
where
$$
B^\mu={d\over dt}\left({\pi^\mu
\over \p2}\right)\qquad.\eqno(49)
$$
We expect that the full component
action of (47) would be equivalent
to a ``double dimensional reduction"
of the action proposed for the d=4
superstring in [9], but we have not checked this.

{\bf 4.} In conclusion, we have succeeded
in developing a $\kappa$-symmetry calculus for the
d=2 and d=3, N=2 massive superparticles.
We first reformulated the superparticle
actions as N=1 and N=2 worldline
supersymmetric sigma models, respectively,
and showed that the $\kappa$-symmetry
is embedded in the local superconformal
symmetry. We then developed a calculus for the latter
without the introduction of a
worldline supergravity multiplet.

The generalisation to the d=5 (or d=4, N=2)
massive superparticle should be
straightforward. Using
harmonic superspace techniques, the
d=6 massless superparticle was reformulated as an N=4
supersymmetric sigma model in [13].
After dimensional reduction we should
obtain a reformulation
of the d=5 massive superparticle as an N=4
supersymmetric sigma model and we should then be able
to develop a calculus for the N=4 local
superconformal symmetry. The generalisation to the d=9
massive superparticle is less obvious.
It is natural to conjecture that it can be reformulated
as an N=8 supersymmetric sigma model,
but straightforward generalisations
run into difficulties because
of the non-associativity of the octonion
algebra. It was suggested in [13] that the notion of
light-like analyticity would be useful.

Perhaps a more interesting question is
the generalisation to higher dimensional supersymmetric
extended objects. Independently, Berkovits
[17,18] and Ivanov and Kapustnikov [15] have reformulated
the superstring action as a supersymmetric
sigma model, but it does involve a partial gauge-fixing of
the superstring action. Furthermore,
Ivanov and Kapustnikov have made
progress towards developing
a $\kappa$-symmetry calculus for the d=4
superstring in this gauge. We hope to
return to further generalisations
in the future.

It was shown in [19], in the context of
massless particles, that the N=1 STV
action in d=3 and the N=2
reformulation for d=4 (different from that
in [13]) could be recast as
Chern-Simons quantum mechanics
actions [20]. The reason why this was
possible is that these models are
invariant under local superconformal
transformations of the worldline without
the introduction of a worldline
supergravity multiplet. As this property
is also shared by our reformulation of
the d=2 and d=3, N=2 massive superparticle
actions and their
higher order generalisations, it should
also be possible to reformulate them
as Chern-Simons quantum mechanics
actions.

We would also like to generalise the
$\kappa$-symmetry calculus to
curved spacetime backgrounds.
To generalise the usual superparticle action,
one has to impose constraints on the
background and we expect that the same
constraints should be sufficient to generalise
the higher order actions.

Finally, it is desirable to have a better
understanding of the relationship
between this $\kappa$-symmetry
calculus and the calculus developed in [10,11,7].
The method of non-linear realisations
was used to construct actions for the d=2 [10,11]
and the d=3, N=2 [7] massive superparticles
with partially broken rigid supersymmetry (PBRS)
and these were interpreted as $\kappa$-invariant
actions
in a particular gauge. Presumably, by gauge-fixing
the local superconformal symmetry, we could solve the
constraints (1) and (29) in terms of
unconstrained superfields
as in the PBRS work.
Alternatively, we expect that the coset methods
used in the PBRS work can be generalised
to derive actions in a manifestly superconformal
and hence $\kappa$-invariant form. For
internal symmetries it is well known that
the non-linear realisation based on
the coset $G/H$ is equivalent to one based
on $G\times H_{local}$ [21] and we
believe that this can be generalised to
include spacetime (super)symmetries.
In this sense it is satisfying to note that
in both formalisms, the massive
superparticle actions themselves emerge as
Wess-Zumino type terms.
In the PBRS approach it is known that this is
related to the presence of central
charges in the spacetime supersymmetry
algebra [11,7] and we expect this also to
be true for the formalism developed in
this paper.
We hope to report on these investigations in the near future.

\bigskip
\bigskip
\bigskip
\bigskip
\leftline{\bf Acknowledgements}

We would like to thank P.G.O. Freund for useful discussions. This work is
supported by a grant from the Math
Discipline Center of the Department of Math, University of Chicago.
\vfill\eject
\leftline{{\bf References}}

\item{
[1]} J. Hughes and J. Polchinski, Nucl.Phys. {\bf B278} (1986) 147.

\item{
[2]} J. Hughes, J. Liu and J. Polchinski, Phys.Lett. {\bf 180B} (1986) 370.

\item{
[3]} P.K. Townsend, Phys.Lett. {\bf 202B} (1988) 53.

\item{
[4]} A. Achucarro, J.P. Gauntlett, K. Itoh and P.K. Townsend, Nucl.Phys. {\bf
B314} (1989) 129.

\item{
[5]} J.P. Gauntlett, Phys.Lett. {\bf 228B} (1989) 188.

\item{[6]} J.A. de Azcarraga, J.P. Gauntlett, J.M. Izquierdo and P.K.
Townsend, Phys.Rev.Lett. {\bf 63} (1989) 2443.

\item{
[7]} J.P. Gauntlett and C.F. Yastremiz, Class. Quantum Grav. {\bf 7} (1990)
2089.

\item{
[8]} E.A. Ivanov and A.A. Kapustnikov, Phys.Lett. {\bf 252B} (1990) 212.

\item{
[9]} T. Curtright and P. van Nieuwenhuizen, Nucl.Phys. {\bf B294} (1987) 125.

\item{
[10]} J.P. Gauntlett, K. Itoh and P.K. Townsend,
Phys.Lett. {\bf 238B} (1990) 65.

\item{
[11]} J.P. Gauntlett, J. Gomis and P.K. Townsend, Phys.Lett. {\bf 249B} (1990)
255.

\item{
[12]} D.P. Sorokin, V.I. Tkach and D.V. Volkov, Mod.Phys.Lett. {\bf A4} (1989)
901.

\item{
[13]} F. Delduc and E. Sokatchev, ``{\it Superparticle with extended worldline
supersymmetry}",
Paris 7 preprint  Par-LPTE/91-14.

\item{
[14]} E.A. Ivanov and A.A. Kapustnikov, ``{\it Gauge covariant Wess-Zumino
actions for super p-branes in superspace}"
Trieste preprint IC/90/425.

\item{
[15]} E. A. Ivanov and A. A. Kapustnikov, ``{\it Towards a tensor calculus for
$\kappa$-symmetry}",
Trieste preprint IC/91/68.

\item{
[16]} P.S. Howe, S. Penati, M. Pernici and P.K. Townsend, Class. Quantum Grav.
{\bf 6} (1989) 1125.

\item{
[17]} N. Berkovits, Phys.Lett. {\bf 232B} (1989) 184.

\item{
[18]} N. Berkovits, Phys.Lett. {\bf 241B} (1990) 497.

\item{
[19]} P.S. Howe and P.K. Townsend, Phys.Lett. {\bf 259B} (1991) 285.

\item{[20]} P.S. Howe and P.K. Townsend, Class. Quantum Grav. {\bf 7} (1990)
1655.

\item{[21]} M. Bando, T. Kugo and K. Yamawaki, Phys.Rep. {\bf 164} (1988) 217.

\end{document}